# Distributed Brillouin frequency shift extraction via a convolutional neural network

Yiqing Chang[1], Hao Wu[1,2], Can Zhao, Li Shen, Songnian Fu and Ming Tang[3]

*Wuhan National Laboratory for Optoelectronics (WNLO) & National Engineering Laboratory for Next Generation Internet Access System, School of Optical and Electronic Information, Huazhong University of Science and Technology, Wuhan 430074, China*

[1]*These authors contribute equally to this work*

[2]*wuhaoboom@qq.com*

[3]*tangming@mail.hust.edu.cn*

**Abstract:** Distributed optical fiber Brillouin sensors detect the temperature and strain along a fiber according to the local Brillouin frequency shift, which is usually calculated by the measured Brillouin spectrum using Lorentzian curve fitting. In addition, cross-correlation, principal component analysis, and machine learning methods have been proposed for the more efficient extraction of Brillouin frequency shifts. However, existing methods only process the Brillouin spectrum individually, ignoring the correlation in the time domain, indicating that there is still room for improvement. Here, we propose and experimentally demonstrate a full convolution neural network to extract the distributed Brillouin frequency shift directly from the measured two-dimensional data. Simulated ideal Brillouin spectrum with various parameters are used to train the network. Both the simulation and experimental results show that the extraction accuracy of the network is better than that of the traditional curve fitting algorithm with a much shorter processing time. This network has good universality and robustness and can effectively improve the performances of existing Brillouin sensors.

## 1. Introduction

Distributed optical fiber sensors can realize a variety of physical quantity measurements at each point along an optical fiber[1]. Among them, distributed optical fiber Brillouin sensors are able to obtain the temperature and strain along a fiber by measuring the distributed Brillouin frequency shift (BFS)[2]. This technology is widely used in the monitoring of large structures, such as bridges and dams, and long-distance temperature measurements for pipelines and tunnels[3]. The distributed BFS is generally obtained by measuring the Brillouin gain spectrum (BGS) of an optical fiber. Since the BGS theoretically satisfies a Lorentzian shape, the BFS can be obtained by performing Lorentzian curve fitting (LCF) on the BGS, which is measured at a limited frequency sampling interval[4]. However, the accuracy of LCF is easily affected by the initial values of the fitting parameters[5,6]. When the signal-to-noise ratio (SNR) is low, improper initial values may lead to a serious error in the fitting result. In addition, the curve fitting algorithm is iterative. Therefore, its processing time is relatively long, which affects the response time of the sensor. To improve the sensing performance, a more accurate and efficient BFS extraction method is needed.

Recently, other methods, such as cross-correlation[5,7], principal component analysis (PCA)[8], and machine learning methods, have been proposed to analyze the BGS[9-13]. Although these algorithms can achieve better results than LCF under certain conditions, they also have some drawbacks. The cross-correlation method calculates the frequency difference to obtain the BFS by convolving the ideal BGS with the measured BGS. It has a higher requirement for the frequency sampling interval than LCF.

Clustering and classification algorithms such as PCA and support vector machine (SVM) have shown good performance for BFS extraction[9-11]. Nevertheless, these algorithms have a trade-off problem between the number of principal components or classes and capability. The accurate extraction of BFS depends on classes or subdivided principal components and large storage databases. Alternatively, artificial neural networks have also been proven effective[12,13]. However, neural networks are usually trained based on specific data. Retraining or fine tuning is required to accommodate different actual data, which significantly affects its application potential. In addition, all of the above methods are designed to analyze only a single BGS at a time to estimate its corresponding BFS. However, the measured result of a distributed Brillouin sensor is natural two-dimensional (2D) data with both time domain and frequency domain information. Existing methods only consider the frequency domain characteristics of the data and do not take advantage of its time domain correlation, indicating that there is still room for improvement[14].

To achieve more efficient BFS extraction with better universality and robustness, a distributed BFS extraction convolutional neural network (CNN) is proposed in this paper. Multilayer 2D convolution is used to analyze the frequency and time features of the measured data and realizes an end-to-end transformation from the 2D data to a one-dimensional (1D) distributed BFS. To adapt the CNN to different instruments and application scenarios, a large number of ideal BGSs with random BFSs, spectral widths (SWs), and SNRs are generated by simulation. These BGSs are randomly combined into 2D data as the input of the CNN, and the corresponding distributed BFS is used as the training target. By optimizing the network structure, training data, and training process, the CNN realizes a high-precision distributed BFS extraction for both the simulation and experiment data. In addition, the proposed network is a full CNN that can fully utilize the parallel computing power of the hardware and effectively reduce the processing time[15].

## 2. Results and discussion

To fully demonstrate the performance of the proposed CNN, first, simulated ideal BGSs with different parameters are generated to verify the universality and robustness of the CNN to various SNRs, BFSs, and SWs. Then, the actual distributed BGS of an optical fiber is measured for testing. These data are processed using the LCF and the trained CNN to compare their performances.

**Simulation results**

**Table 1 Test Parameters**

| Parameters | Range | Interval |
|---|---|---|
| SNR | 5~19 dB | 2 dB |
| BFS | 10%~90% | 5% |
| SW | 10%~50% | 5% |

First, the performance of the CNN at different SNRs is analyzed. The SW of the simulated BGSs is fixed at 25%, and the BFS is fixed at 30% (normalized by the frequency sampling range; see the data preparation subsection in the Materials and Methods section). For each SNR in Table 1, 4480 noisy BGSs are generated by adding Gaussian white noise. The root-mean-square error (RMSE) and standard deviation (SD) are used to evaluate the BFS extracted by the LCF and CNN (see the performance evaluation subsection in Materials and Methods section). As shown in Fig. 1(a) and (b), the RMSE and

SD decrease as the SNR increases. The RMSE of the BFS extracted by the CNN is better than that extracted by the LCF when the SNR is lower than 16 dB. A similar trend is found in the SD when the SNR is lower than 18 dB. By utilizing the 2D information of the BGSs, the CNN can extract the BFS from the noisy data more accurately. Since the simulated BGSs are generated based on the Lorentzian shape, the LCF can achieve better results when the amount of noise is low.

To show the effects of the CNN on different BFS data, the SW and SNR are chosen to be 25% and 11 dB, respectively. For each BFS in Table 1, 4480 BGSs are simulated for testing. As shown in Fig. 1(c) and (d), the processing results of the CNN are better than those of the LCF for all cases. In addition, the CNN has good consistency for different BFS data. However, when the BFS approaches the edge of the scanning range, the results using the LCF significantly deteriorate. This means that the CNN has a higher tolerance for the incompleteness of the BGS and can achieve high-precision BFS extraction with less information.

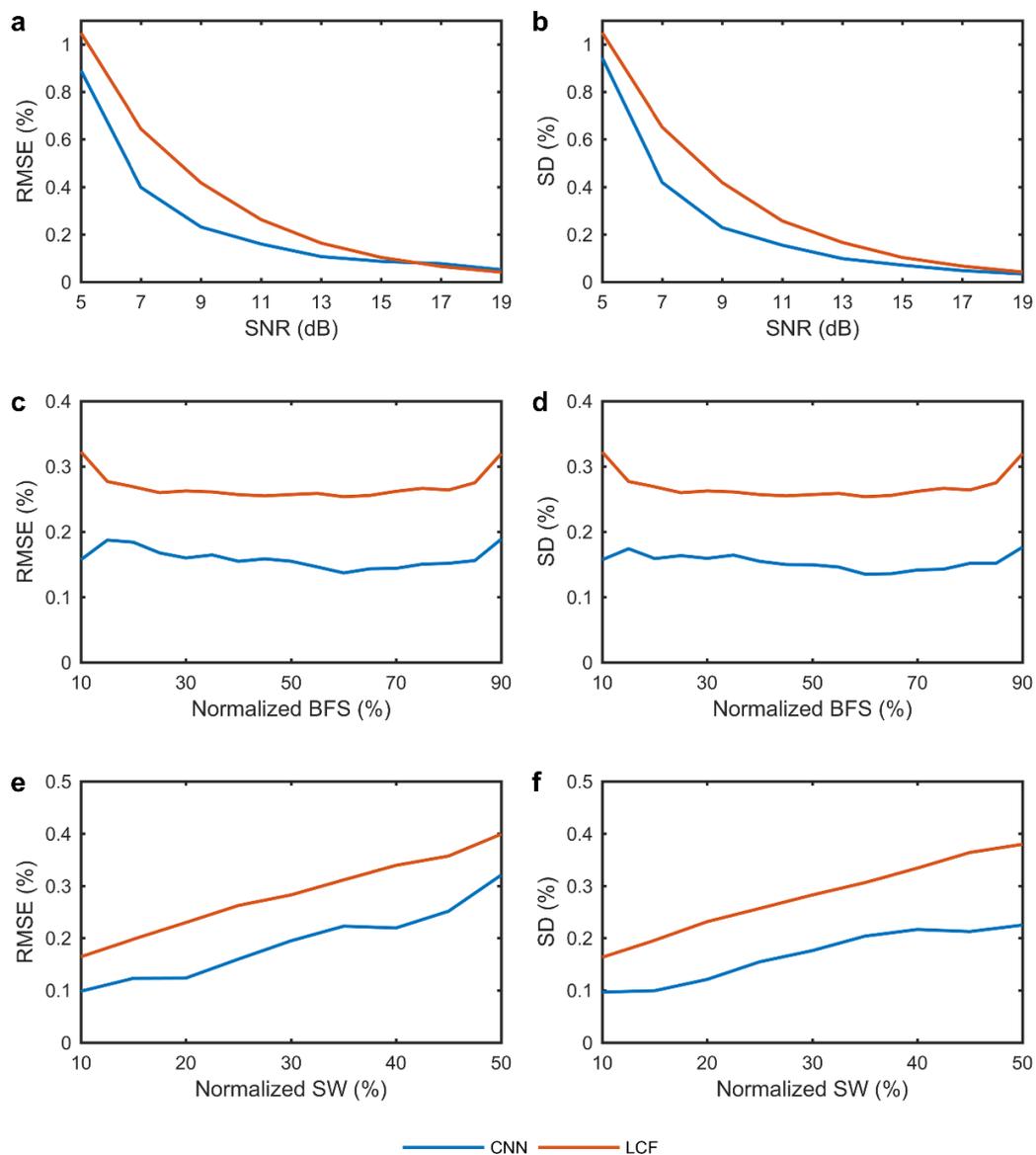

**Fig. 1 Normalized BFS RMSE and SD of the simulation data.** Normalized RMSE **a** and SD **b** for different SNR data. Normalized RMSE **c** and SD **d** for different BFS data. Normalized RMSE **e** and SD **f** for different SW data

BGSs with various SWs are also simulated, while the SNR and BFS are fixed at 11 dB and 25%, respectively. As shown in Fig. 1(e) and (f), it is harder to extract the BFS accurately via both methods when the SW increases. The results of the CNN are always better than those of the LCF. For the LCF results, the changes in RMSE and SD with SW are basically consistent. However, the results obtained by the CNN have less deterioration in the SD when the SW is large. Because the CNN exploits the time domain characteristics of the data, there is a correlation between adjacent BGSs, which causes the SD to be small.

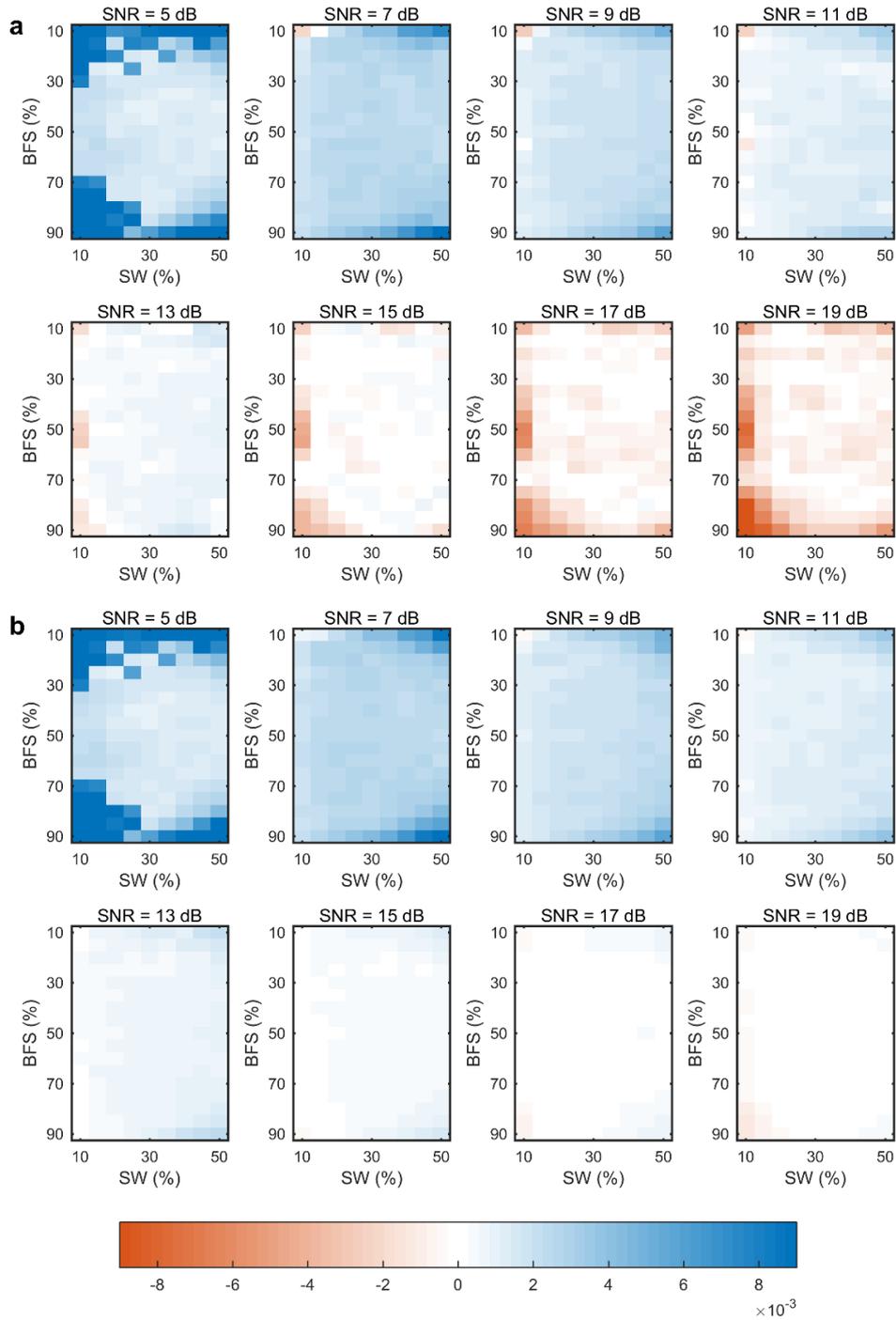

Fig. 2 Performance differences in BFSs extracted by the LCF and CNN. a Normalized BFS RMSE using the CNN minus the normalized BFS RMSE using the CNN for different simulation data. b Normalized BFS SD using the CNN minus the normalized BFS SD using the CNN for different simulation data.

To compare the performances of the CNN and LCF more comprehensively, 4480 BGSs are simulated for each case, as shown in Table 1. We subtract the BFS RMSE using the CNN from that of the result using the LCF and plot the differences in Fig. 2(a). Analogously, Fig. 2(b) shows the SD differences. Positive results are shown in blue, indicating that the RMSE or SD using the LCF is larger than that using the CNN. Red indicates that the CNN performs worse than the LCF in that case. In addition, the images with darker tones represent larger performance differences. The results indicate once again that the CNN can extract the BFS more accurately when the SNR is low, and the correlation between different BGSs is stronger under this algorithm than the LCF. In the cases of low SNR, the values near the edges in the subgraphs are smaller, which indicates that the CNN is more robust to the BGS than the LCF and can handle more extreme cases.

**Experimental results**

To demonstrate the validation of the trained CNN on the actual distributed Brillouin sensors, a Brillouin optical time domain analyzer (BOTDA) system is set up to measure the distributed BGS data of a stand single-mode fiber that is approximately 25 km long. The time domain sampling rate is 250 MSa/s, and the frequency scanning range is from 10.6 GHz to 10.9 GHz at a step of 2 MHz. Figure 3(a) shows the measured distributed BGSs when the pump pulse width is 40 ns and the average time is 32. This distributed BGS is processed by the CNN and LCF, and the extracted BFSs are shown in Fig. 3(b). The values of the extracted BFSs are almost the same with some fluctuation. As the SNR decreases with distance, the fluctuation range becomes more severe. However, the fluctuation is obviously weaker when using the CNN than when using the LCF. However, the magnitude of the fluctuation is generally not used directly to judge the accuracy of extraction. Because the true BFS of the fiber is unknown, the fluctuation may be caused by temperature and strain.

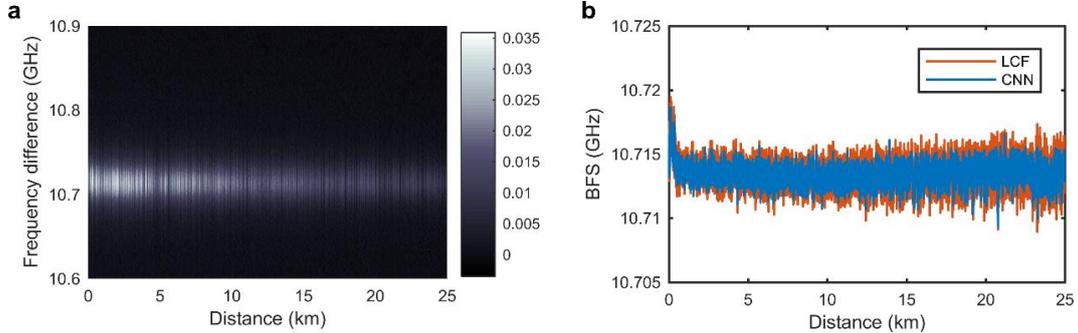

**Fig. 3 Measurement results. a** Measured BGSs along an optical fiber. **b** Distributed BFSs extracted by an LCF and CNN

Here, we use uncertainty as the basis for evaluating the performance of the extraction algorithms[4]. The distributed BGSs of the same fiber are continuously measured, and the data are processed by the CNN and LCF. The uncertainty is defined as the quadratic fitted trace of the SD of the continuous extracted BFSs. To illustrate the universality and robustness of the proposed CNN, several experiments are performed with different average times and pulse widths. Average times of 8 and 32 and pulse widths of 20 ns, 30 ns, and 40 ns are selected to achieve a large SNR range and SW variation. As shown in Fig. 4, the CNN performs better than the LCF in all cases. When the uncertainty reaches below 0.4 MHz, the CNN still has certain advantages, which verifies the conjecture in the simulation section: the CNN may work better than the LCF for actual BGS when the SNR is high.

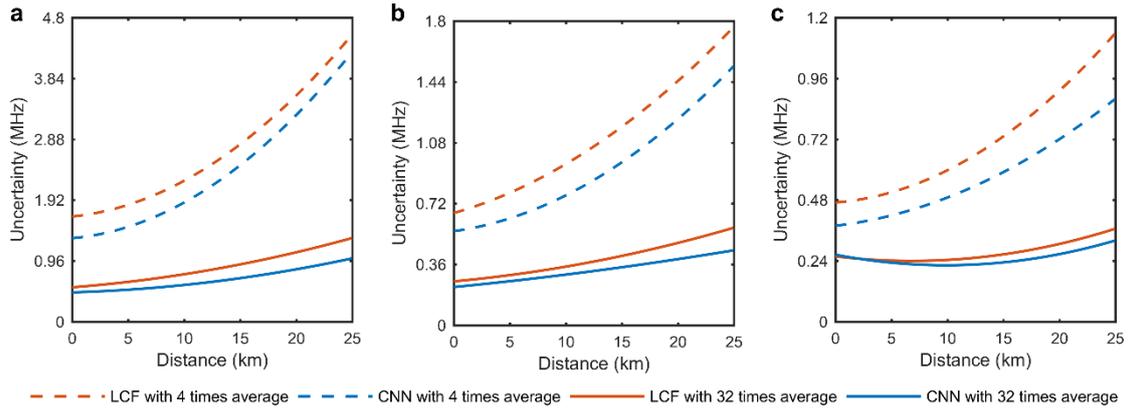

**Fig. 4 BFS uncertainty as a function of fiber length.** BFS uncertainty traces when the pump pulse width is 20 ns **a**, 30 ns **b**, and 40 ns **c**.

**Spatial resolution**

The spatial resolution is a key parameter for distributed Brillouin sensors and is defined as the fiber length of the BFS transition region between 10% and 90% of the peak frequency. To investigate whether the spatial resolution is affected by the CNN, approximately 100 m of the fiber end is placed in a temperature-controlled chamber (TCC) and heated to 50 °C, while the rest of the fiber is at room temperature. Figure 5 shows the BFSs extracted by the LCF and CNN when the pump pulse width is 40 ns and the average time is 32. The BFSs obtained by the two methods are basically the same in the transition region. Because the BFS of the training data varies randomly, the CNN can adapt to arbitrary data without changing the spatial resolution.

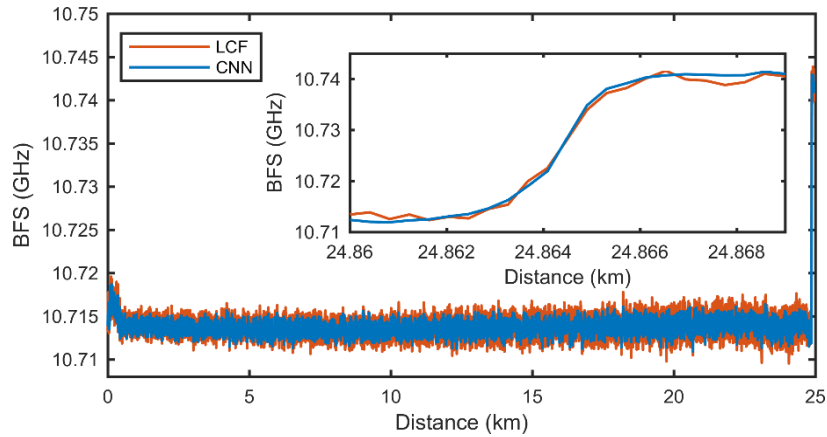

**Fig. 5 Extracted BFSs along the fiber when the last 100 m of the fiber is heated.** The inset image shows the BFS profiles around the start of the heat section.

**Processing time**

For distributed Brillouin sensors, data processing time is one of the key factors affecting the system response time. Especially for high-speed acquisition systems, the LCF has become a bottleneck[16,17]. It takes approximately 0.129 s for the CNN to process 1000 BGSs based on a Python environment running on a PC with an AMD Ryzen 7 1700 eight-core processor and a Nvidia GeForce GTX 1070 Ti (8 GB) GPU. For the same data and operating environment, the LCF takes approximately 0.814 s, which is much longer than the CNN. It should be pointed out that the results of previous articles are generally running on MATLAB software. Therefore, the LCF processing time for 1000 BGSs using MATLAB is also given as a reference here, which is approximately 12.14 s.

## 3. Conclusions

In this paper, a CNN is employed for the BFS extraction of distributed Brillouin sensors. Because of the full convolutional structure, the CNN can obtain the distributed BFS directly from the measured 2D distributed BGS. By making full use of the information of the 2D data, the CNN achieves better BFS extraction accuracy than the traditional LCF. Both the simulation and experimental results confirm that the CNN has better universality and robustness. Due to the comprehensive analysis of adjacent BGSs, the uncertainty of the BFS obtained by the CNN has been significantly improved. In addition, the processing time of the CNN is much faster thanks to its suitable net structure for parallel computation.

The proposed CNN can be applied to any distributed Brillouin sensor and effectively improve the performance of the sensor with no hardware modification.

## 4. Materials and methods

### Lorentzian curve fitting

The most traditional method to obtain the BFS is the LCF:

$$g(v) = \frac{g_B}{1 + \left(\frac{v - v_B}{\Delta v_B / 2}\right)^2} \quad (1)$$

where $g_B$ is the Brillouin gain coefficient; $\Delta v_B$ is the full width at half maximum of the spectrum, which is the SW; and $v_B$ represents the BFS. Before fitting, the initial values of these parameters should be set. The maximum value of the BGS is assigned to $g_B$, and its corresponding frequency is assigned to $v_B$. The initial value of the SW is estimated by calculating the frequency range where the intensity exceeds half of the maximum gain.

Starting with the initial values, the least squares method is used to find parameters to best fit equation (1) iteratively. That is, for a set of measured data ($v_i$, $g_i$), the purpose of the method is to find the parameters $\hat{a}$ so that the sum of the square error is minimized:

$$\hat{a} \equiv \operatorname{argmin}_a \sum_{i=1}^{m} \left[g_i - g(v_i; a)\right]^2 \quad (2)$$

The Levenberg-Marquardt algorithm (LMA) is employed to solve this nonlinear least squares problem[6,18]. The LMA combines the advantages of the Gauss-Newton algorithm and gradient descent method to obtain better robustness. Based on the LMA, the parameters are continuously evolved through iteration until the iteration step is less than the stopping criteria, which is set to $10^{-8}$.

### Neural network architecture

As illustrated in Fig. 6, the proposed network consists of three parts. The first part starts with an input layer of $151 \times N \times 1$. The number 151 represents the number of frequency sampling points, which is a general choice considering the sampling time. N means the number of input BGS traces. It is important to note that there is a one-to-one correspondence between the input BGS and the output BFS. Therefore, the length of the signal at the time dimension should not be reduced over the network during the process. The input data are first processed by 64 convolutional filters of size $3 \times 3$ to generate 64 feature maps. After that, a max pooling layer, with a spatial extent of $2 \times 1$ and a step size of $2 \times 1$, is used to reduce the size of the feature map to allow the CNN to focus on more valuable information. The second part is an

18-layer residual subnetwork[19]. There are 6 ResBlocks with the structure of Conv (1×1) – Conv (3×3) – Conv (1×1). The 3×3 convolution kernels are used to perceive futures in both the time and frequency domains. The 1×1 convolution kernels are employed to be introduced more nonlinearly by using a rectified linear unit (RELU). All convolutional layers in the first two parts are in same-padding mode. In addition, batch normalization (BN) is adopted after each convolution layer and before activation. The third part is a plain net that aims to obtain the 1D BFS. Therefore, M×1 convolution kernels are used. After testing and optimization, a total of 7 layer schemes that utilize 7×1 and 3×1 convolution kernels are chosen. Finally, the output size is 1×N, corresponding to the BFSs for N input BGSs.

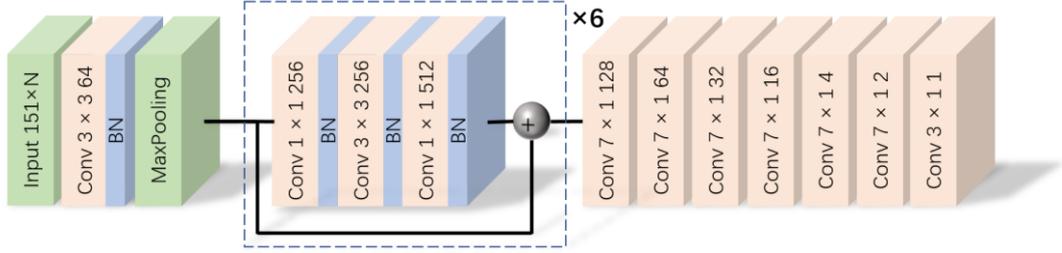

**Fig. 6 Architecture of the proposed CNN.**

### Data preparation

To ensure the efficient operation of the network, the input data must be preprocessed, as shown in Fig. 7. First, the maximum value of the BGS is transformed to 1. The BFS is normalized according to the sweep range:

$$BFS_N = \frac{BFS - f_{min}}{f_{max} - f_{min}} \quad (3)$$

where $f_{max}$ and $f_{min}$ represent the maximum and minimum values of the sweep range, respectively. The SW is normalized according to the sweep range in the same way and expressed as a percentage.

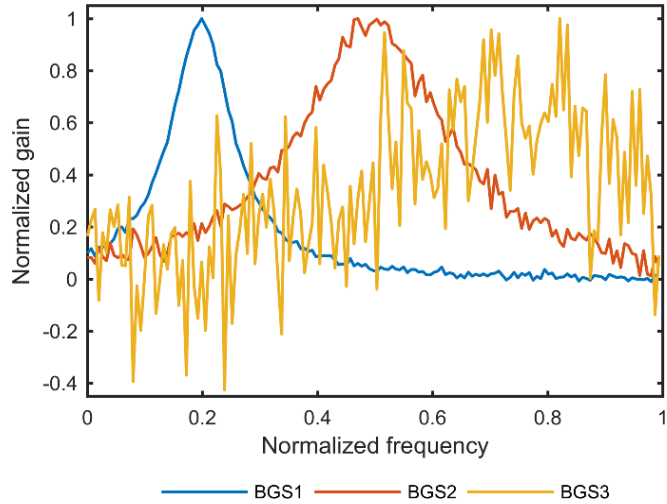

**Fig. 7 Normalized simulation BGSs.** BGS1: BFS=20%, SW=13%, SNR=20 dB; BGS2: BFS=50%, SW=30%, SNR=15 dB; BGS3: BFS=73%, SW=50%, SNR=10 dB.

### Training process

Simulated BGSs are used to train the proposed CNN. The BGSs are generated as Lorentzian curves with a random BFS and SW. To further enhance the robustness, Gaussian white noise is added to the ideal BGSs with a random SNR. The SNR is defined as the ratio between the maximum possible power of a signal and the power of noise. Considering the actual Brillouin system, the random range of the BFS is set to 5% to 95%, the range of the SW is 10% to 50%, and the range of the SNR is 5 to 20 dB. By combining the simulated BGSs randomly, 2D data are generated as the input of the CNN, and its corresponding distributed BFS is used as the training target, as shown in Fig. 8.

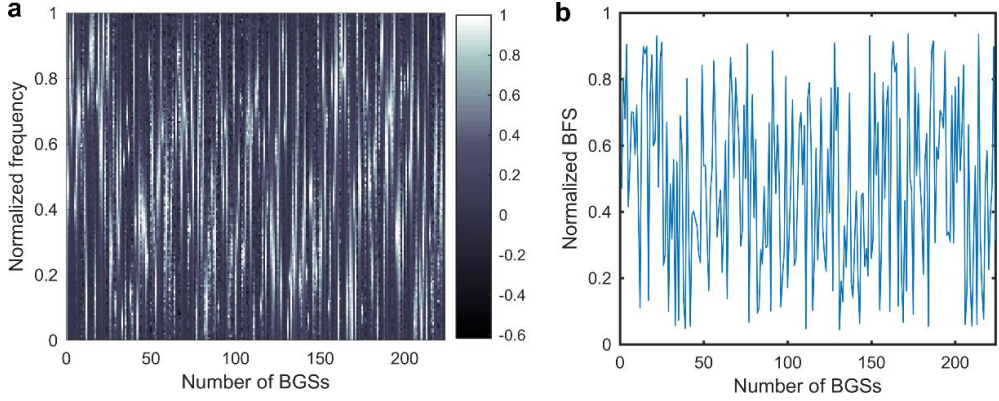

**Fig. 8 Simulation data used for training. a** Simulation BGSs of random parameters. **b** Corresponding BFSs

First, the parameters of the CNN are initialized randomly as a uniform distribution. Then, the optimization function Adam is employed to optimize the parameters according to the training data[20]. For each iteration, the training data go through forward propagation, as shown in equation (4). Then, the backward propagation of the loss value is calculated according to equation (5). Finally, the network parameters are updated based on equation (6).

$$x^l = \begin{cases} 1, & l = 1 \\ \sigma(w^l x^{l-1} + b^l), & l = 2 \text{ to } L \end{cases} \quad (4)$$

$$\delta^l = \begin{cases} \text{MSE}(z^l), & l = L \\ \delta^{l+1} * \text{rot}180(w^{l+1}) \odot \sigma'(z^l), & l = L-1 \text{ to } 2 \end{cases} \quad (5)$$

$$\theta_{t+1}^l = \theta_t^l - \alpha \frac{\hat{m}_t}{\sqrt{\hat{v}_t + \varepsilon}} \quad (6)$$

where $x^l$ represents the output of the $l$ layer in the forward propagation and $\delta^l$ represents the gradient of the $l$ layer in backward propagation. $\theta^l$ refers to the parameters of the $l$ layer, including weight $w$ and bias $b$. $\sigma$ and MSE are the activation function and loss function, respectively. The symbol $\odot$ in equation (5) represents the Hadamard product. In equation (6), $\hat{m}_t$ and $\hat{v}_t$ are bias-corrected first moment and second raw estimations in Adam, respectively. In addition, $a$ is the learning rate.

The model is trained for 22 epochs with a mini-batch size of 8. For each training epoch, the CNN is updated 375 times using 672000 BGSs. The learning rate starts from 0.001 and gradually decreases with a decay of 0.0001 over each update. It takes approximately 2 hours and 53 minutes to complete the training process on the same environment for testing.

**Experimental setup**

To measure the actual distributed BGS of an optical fiber, a BOTDA system is set up, as shown in Fig. 9. The output of a narrow linewidth laser at 1550 nm is split into two branches through a 3 dB coupler. The probe light in the upper branch is modulated by an EOM operated in carrier-suppression mode to produce sidebands. The EOM is driven by an MS to control the frequency of the sidebands. Then, the probe passes through a PS to mitigate the polarization effects and launches into one end of the fiber.

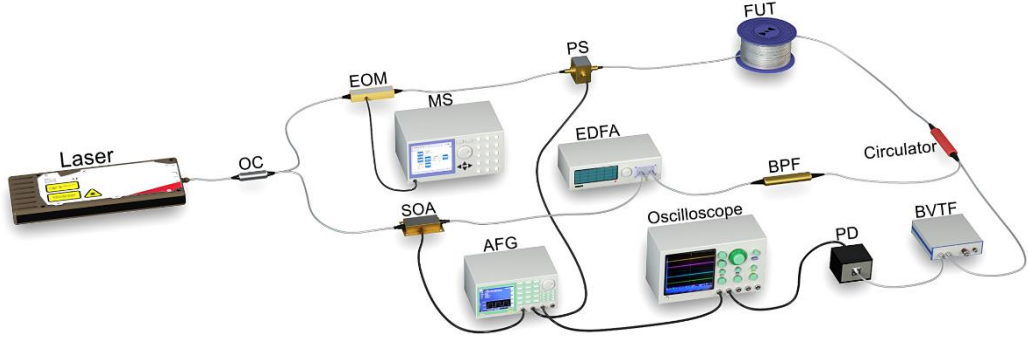

**Fig. 9 Experimental setup of the BOTDA system.** EOM electro-optic modulator, MS microwave synthesizer, PS polarization switch, TCC temperature-controlled chamber, SOA semiconductor optical amplifier, AFG arbitrary function generator, EDFA erbium-doped fiber amplifier, BPF bandpass filter, BVTF bandwidth-variable tunable filter, PD pin photodetector

In the lower branch, an SOA driven by an AFG is exploited to generate optical pump pulses with a high extinction ratio (>50 dB). After amplification by an EDFA, the pump pulses are launched into the other end of the fiber through a circulator. Due to the stimulated Brillouin scattering effect, a part of the energy of the high-frequency pump light is transferred to the low-frequency probe light. Finally, the Brillouin-amplified probe wave is obtained through a BVTF and eventually detected by a 125 MHz PD. By continuously sweeping the frequency difference between the pump and probe light around the local BFS, the distributed BGS is obtained.

**Performance evaluation**

The RMSE and SD are used as evaluation parameters to compare the performances of the LCF and CNN. As shown in equation (7), the RMSE characterizes the difference between the extracted BFS and its true value. The SD characterizes the dispersion of the extracted BFS.

$$RMSE = \sqrt{\frac{1}{N}\sum_{i=1}^{N}(y-\hat{y})^2} \tag{7}$$

$$SD = \sqrt{\frac{1}{N}\sum_{i=1}^{N}(y-\bar{y})^2} \tag{8}$$

where y and ŷ are the predicted and true BFSs, respectively, and $\bar{y}$ is the average of the predicted BFSs.

## 5. Acknowledgements

This work is supported by the National Key R&D Program of China (2018YFB1801002), National Natural Science Foundation of China (NSFC) (61722108, 61931010), and Innovation Fund of WNLO.